\documentstyle[epsfig]{aipproc}

\begin{document}
\title{Infinite Dimensional Quantum Information Geometry}
\author{Matheus R. Grasselli\thanks{Supported by a grant from
Capes-Brazil.}}
\address{King's College London \\Strand,\\
London, WC2R 2LS}
\maketitle
\newtheorem{corollary}[equation]{Corollary}
\newtheorem{theorem}[equation]{Theorem}
\newtheorem{lemma}[equation]{Lemma}
\newtheorem{definition}[equation]{Definition}
\renewcommand{\theenumi}{\roman{enumi}}

\begin{abstract}
We present the construction of an infinite dimensional Banach
manifold of quantum mechanical states on a Hilbert space
$\mathcal{H}$ using different types of small perturbations of a
given Hamiltonian $H_0$. We provide the manifold with a flat
connection, called the exponential connection, and comment on the
possibility of introducing the dual mixture connection
\end{abstract}

\section*{Introduction}

In finite dimensional quantum information geometry, the set upon
which the geometric structures are defined is simply the set of
all (invertible) density matrices on a finite dimensional Hilbert
space \cite{grasselli:PetzSudar96}. Already in the definition of
the underlying set in infinite dimensions, we need to be slightly
more careful and take a more restrictive set than just that of all
(invertible) density operators. The reason for this is that, if we
modify a given state in the set with a small perturbation, we want
the perturbed state to have the same properties as the original
one.

Let ${\cal C}_p,0<p<1$, denote the set of compact operators
$A:\cal H \mapsto \cal H$ such that $|A|^p \in {\cal C}_1$, where
${\cal C}_1$ is the set of trace-class operators on $\cal H$.
Define ${\cal C}_{<1}:=\bigcup_{0<p<1} {\cal C}_p$. We take the
underlying set of the quantum information manifold to be ${\cal
M=C}_{<1}\cap\Sigma$ where $\Sigma \subseteq {\cal C}_1$ denotes
the set of density operators. This guarantees that, if $\rho_0 \in
\mathcal{M}$, there exists a $\beta_0<1$ such that
$\rho_0^{\beta_0}$ is a density as well as $\rho_0$ itself. This
set has an affine structure induced from the linear structure of
each ${\cal C}_p$ in the following way: let $\rho_1 \in {\cal
C}_{p_1}\cap \Sigma$ and $\rho_2 \in {\cal C}_{p_2}\cap \Sigma$;
take $p=\max \{p_1,p_2\}$, then $\rho_1,\rho_2 \in {\cal
C}_{p}\cap \Sigma$, since $p\leq q$ implies ${\cal C}_p \subseteq
{\cal C}_q$ \cite{grasselli:Pietsch72}; define
``$\lambda\rho_1+(1-\lambda)\rho_2, 0 \leq \lambda \leq 1$'' as
the usual sum of operators in ${\cal C}_p$. This is called the
$(-1)$-affine structure. However, this is not the affine structure
we use to define a flat connection on the manifold. Instead, we
equip ${\cal M}$ with an exponential affine structure and then
define the natural connection associated with it.

To each $\rho_0 \in {\cal C}_{\beta_0}\cap \Sigma$, $\beta_0 < 1$,
let $H_0=-\log \rho_0 + cI \geq I$ be a self-adjoint operator with
domain ${\cal D}(H_0)$ such  that
$\rho_0=Z_0^{-1}e^{-H_0}=e^{-(H_0+\Psi_0)}$.

The idea is to perturb the Hamiltonian $H_0$, obtaining a new
Hamiltonian $H_X$ and then construct a neighbourhood of the point
$\rho_0$ consisting of the perturbed states $\rho_X$. The
perturbations considered are of three different types, and that is
the content of the next section.

\section*{Perturbations}

The most general class of perturbations we use are the form
bounded perturbations. Given a positive self-adjoint operator $H$
with associated form $q_H$ and form domain $Q(H)$, we say that a
symmetric quadratic form $X$ (or the symmetric sesquiform obtained
from it by polarization) is {\em $q_H$-bounded\/} if
\begin{enumerate}
  \item $Q(H)\subset Q(X)$ and
  \item there exist positive constants $a$ and $b$ such that
  $\left|X(\psi,\psi)\right|\leq aq_H(\psi,\psi) +
  b(\psi,\psi)$, for all $\psi\in Q(H)$.
\end{enumerate}

Although form bounded perturbations are of much interest in the
study of Schr\"{o}dinger operators for a variety of quantum
systems, they provide very little regularity for the quantum
information manifold. In \cite{grasselli:Streater98}, Streater was
able to show that the manifold constructed using then has
Lipschitz structure. However, more regularity is needed if we want
to define a metric on it by, say, the second derivative of the
free energy. This led to the idea of looking at the more
restrictive case of operator bounded perturbations. Given
operators $H$ and $X$ defined on dense domains ${\cal D}(H)$ and
${\cal D}(X)$ in a Hilbert space $\cal H$, we say that $X$ is {\em
$H$-bounded\/} if
\begin{enumerate}
  \item ${\cal D}(H)\subset{\cal D}(X)$ and
  \item there exist positive constants $a$ and $b$ such that
  $\left\|X\psi\right\|\leq a\left\|H\psi\right\| +
  b\left\|\psi\right\|$, for all $\psi\in{\cal
  D}(H)$.
\end{enumerate}
For both form bounded and operator bounded perturbations, the
infimum of such $a$ is called the relative bound of $X$ (with
respect to $H$ or with respect to $q_H$, accordingly). If $a<1$,
the perturbation is said to be small.

Operator bounded perturbations are also used in the study of
Schr\"{o}dinger operators for quite a broad range of quantum
systems \cite{grasselli:ReedSimon75}, with the additional property
of providing enough regularity for the manifold ${\cal M}$ to have
an analytic free energy, for instance.

The following lemma tells us how to characterise operator bounded
and form bounded perturbations in terms of certain norms. Before
we state it, we need to say what we mean by a form multiplied from
both sides by operators. Suppose that $X$ is a quadratic form with
domain $Q(X)$ and $A,B$ are operators on $\cal H$ such that $A^*$
and $B$ are densely defined. Suppose further that $A^*:{\cal
D}(A^*)\rightarrow Q(X)$ and $B:{\cal D}(B)\rightarrow Q(X)$. Then
the expression $AXB$ means the form defined by
\[\phi,\psi\mapsto X(A^*\phi,B\psi),\qquad \phi\in {\cal
D}(A^*),\quad \psi\in {\cal D}(B).\] Consider now the case where
$H_0\geq I$ is a self-adjoint operator with domain ${\cal
D}(H_0)$, quadratic form $q_0$ and form domain $Q_0={\cal
D}(H_0^{1/2})$, and let $R_0=H_0^{-1}$ be its resolvent at the
origin.

\begin{lemma}
A symmetric operator $X:{\cal D}(H_0)\rightarrow \cal H$ is
$H_0$-bounded if and only if $\left\|XR_0\right\|<\infty$.
Analogously, a symmetric quadratic form $X$ defined on $Q_0$ is
$q_0$-bounded if and only if $R_0^{1/2}XR_0^{1/2}$ is a bounded
symmetric form defined everywhere. Moreover, if
$\left\|R_0^{1/2}XR_0^{1/2}\right\|<\infty$ then the relative
bound $a$ of $X$ with respect to $q_0$ satisfies $a\leq
\left\|R_0^{1/2}XR_0^{1/2}\right\|$.
\end{lemma}

The set ${\cal T}_{\omega}(0)$ of all $H_0$-bounded symmetric
operators X is a Banach space with norm $\|X\|_\omega (0)
:=\left\|XR_0\right\|$, since the map $A\mapsto AH_0$ from $\cal
B(H)$ onto ${\cal T}_{\omega}(0)$ is an isometry. The set ${\cal
T}_0(0)$ of all $q_0$-bounded symmetric forms $X$ is also a Banach
space with norm $\|X \|_0
(0):=\left\|R_0^{1/2}XR_0^{1/2}\right\|$, since the map $A\mapsto
H_0^{1/2}AH_0^{1/2}$ from the set of all bounded self-adjoint
operators on $\cal H$ onto ${\cal T}_0(0)$ is again an isometry.

Motivated by Banach space interpolation theory, let us consider,
for $\varepsilon\in (0,1/2)$, the set ${\cal T}_{\varepsilon}(0)$
of all symmetric forms X with ${\cal
D}(H_0^{\frac{1}{2}-\varepsilon}) \subset Q(X)$ and such that $\|X
\|_\varepsilon
(0):=\left\|R_0^{\frac{1}{2}+\varepsilon}XR_0^{\frac{1}{2}-\varepsilon}\right\|$
is finite. Then the map $A\mapsto
H_0^{\frac{1}{2}-\varepsilon}AH_0^{\frac{1}{2}+\varepsilon}$ is an
isometry from the set of all bounded self-adjoint operators on
$\cal H$ onto ${\cal T}_{\varepsilon}(0)$. Hence ${\cal
T}_{\varepsilon}(0)$ is a Banach space with the $\varepsilon$-norm
$\|\cdot\|_\varepsilon (0)$. Such an $X$ will be called a
$\varepsilon$-bounded perturbation and it is shown to interpolate
between the extreme cases of form bounded perturbations, for which
$\varepsilon=0$, and operator bounded perturbations for which
$\varepsilon=1/2$.

\begin{lemma}
For fixed symmetric $X$, $\left\|X\right\|_\varepsilon$ is a
monotonically increasing function of $\varepsilon\in [0,1/2]$.
\label{monotonicity}
\end{lemma}

In the next section, we carry out the programme of using these
perturbations to obtain the hoods in the manifold. In what
follows, we write ${\cal T}_{(\cdot)}(0)$ to indicate that we can
use any of the three Banach spaces ${\cal T}_{\omega}(0)$, ${\cal
T}_0(0)$ or ${\cal T}_{\varepsilon}(0)$.

\section*{Construction of the Manifold}

The two technical tools used in the construction of our manifold
are the following.

\begin{theorem}[KLMN]
Let $H_0$ be a positive self-adjoint
operator with quadratic form $q_0$ and form domain $Q_0$; let $X$
be a $q_0$-small symmetric quadratic form. Then there exists a
unique self-adjoint operator $H_X$ with form domain $Q_0$ such
that \[\langle H^{1/2}_X\phi,H^{1/2}_X\psi\rangle = q_0(\phi
,\psi)+X(\phi ,\psi),\qquad \phi ,\psi\in Q_0.\] Moreover, $H_X$
is bounded below by $-b$.
\end{theorem}

\begin{lemma}[Streater 98]
Let $X$ be a $q_0$-small with bound $a<1-\beta_0$. Denote  by
$H_X$ the unique operator given by the KLMN theorem. Then $\exp
(-\beta H_X)$ is of trace class for all
$\beta>\beta_X=\beta_0/(1-a)$. \label{traceclass}
\end{lemma}

The construction of the neighbourhood of $\rho_0$ goes as follows.
In ${\cal T}_{(\cdot)}(0)$, take $X$ such that $\|X\|_{(\cdot}(0)
< 1-\beta_0$. Since $\|X \|_0 (0) \leq \|X \|_{(\cdot)}(0) <
1-\beta_0$, $X$ is also $q_0$-bounded with bound $a_0$ less than
$1-\beta_0$. The {\em KLMN} theorem then tells us that there
exists a unique semi-bounded self-adjoint operator $H_X$ with form
$q_X = q_0 + X$ and form domain $Q_X = Q_0$. We write $H_X=H_0+X$
for this operator and consider the state $\rho_X =
Z_X^{-1}e^{-(H_0+X)}=Z_X^{-1}e^{-(H_0+X+\Psi_X)}$

Then, from lemma~\ref{traceclass}, $\rho_X \in {\cal
C}_{\beta_X}\cap \Sigma$, where $\beta_X =
\frac{\beta_0}{1-a_0}<1$. If we add to $H_X$ a multiple of the
identity, we can still have the same state $\rho_X$, by simply
adjusting the partition function $Z_X$; so we can always assume
that, for the perturbed state, we have $H_X \geq I$. We take as a
hood ${\cal M}_0$ of $\rho_0$ the set of all such states, that is,
${\cal M}_0=\{\rho_X:\|X \|_{(\cdot)} (0) <1-\beta_0\}$.

To give a topology to ${\cal M}_0$, we first introduce in ${\cal
T}_{(\cdot)}(0)$ the equivalence relation \mbox{$X \sim Y$} iff
$X-Y=\alpha I$ for some $\alpha \in {\mathbf R}$, precisely
because $\rho_X = \rho_{X+\alpha I}$, as remarked above. We then
identify $\rho_X$ in ${\cal M}_0$ with the line $\{Y \in {\cal
T}_{(\cdot)}(0): Y=X+\alpha I, \alpha \in {\mathbf R}\}$ in ${\cal
T}_{(\cdot)}(0)/\! \sim$. This is a bijection from ${\cal M}_0$
onto the subset of ${\cal T}_{(\cdot)}(0)/\! \sim$ defined by
$\left\{ \{X+\alpha I\}_{\alpha \in {\mathbf R}}:\|X
\|_{(\cdot)}(0)< 1-\beta_0\right\}$. The topology in ${\cal M}_0$
is then given by transfer of structure. Now that ${\cal M}_0$ is a
(Hausdorff) topological space, we want to parametrise it by an
open set in a Banach space. As in the classical case
\cite{grasselli:PistoneSempi95}, we choose the Banach subspace of
centred variables in ${\cal T}_{(\cdot)}(0)$; in our terms,
perturbations with zero mean (the `scores'). The only problem is
that, when $X$ is not an operator, it is not immediately clear
what its mean in the state $\rho_0$ should be, let alone the
question of whether or not it is finite.  To deal with this,
define the regularised mean of $X \in {\cal T}_{(\cdot)}(0)$ in
the state $\rho_0$ as $\rho_0\! \cdot \! X
:=\mbox{Tr}(\rho_0^\lambda X \rho_0^{1-\lambda})$, for $0<\lambda
<1$.

Since $\rho_0 \in {\cal C}_{\beta_0}\cap \Sigma$ and $X$ is
$q_0$-bounded, lemma 5 of \cite{grasselli:Streater98} ensures that
$\rho_0 \! \cdot \! X$ is finite and independent of $\lambda$. It
was a shown there that $\rho_0 \! \cdot \! X$ is a continuous map
from ${\cal T}_0 (0)$ to ${\mathbf R}$, because its bound
contained a factor $\|X \|_0 (0)$. Exactly the same proof shows
that $\rho_0 \! \cdot \! X$ is a continuous map from ${\cal
T}_{(\cdot)}(0)$ to ${\mathbf R}$. Thus the set $\widehat{\cal
T}_{(\cdot)}(0):=\left\{X \in {\cal T}_{(\cdot)} (0):\rho_0 \!
\cdot \! X=0 \right\}$ is a closed subspace of ${\cal T}_{(\cdot)}
(0)$ and so is a Banach space with the norm $\left\| \cdot
\right\|_\varepsilon$ restricted to it. We notice that for the
case of operator bounded perturbations, the regularised mean of
$X$ coincides with the usual mean $\mbox{Tr}(\rho_0X)$.

To each $\rho_X \in {\cal M}_0$, consider the point in the line
$\{X+\alpha I \}_{\alpha \in {\mathbf R}}$ with $\alpha=-\rho_0 \!
\cdot \! X$. Write $\widehat{X}=X-\rho_0 \! \cdot \! X$ for this
point. The map $\rho_X \mapsto \widehat{X}$ is a homeomorphism
between ${\cal M}_0$ and the open subset of $\widehat{\cal
T}_{(\cdot)} (0)$ defined by $\left\{\widehat{X}:
\widehat{X}=X-\rho_0 \! \cdot \! X,
\left\|X\right\|_{(\cdot)}<1-\beta_0 \right\}$. The map $\rho_X
\mapsto \widehat{X}$ is then a global chart for the Banach
manifold ${\cal M}_0$ modeled by $\widehat{\cal T}_{(\cdot)}(0)$.
The tangent space at $\rho_0$ is given by $\widehat{\cal
T}_{(\cdot)}(0)$, with the curve $\left\{\rho (\lambda)=Z_{\lambda
X}^{-1}e^{-(H_0+\lambda X)}, \lambda \in [-\delta
,\delta]\right\}$ having tangent vector $\widehat{X}=X-\rho_0 \!
\cdot \! X$.

We extend our manifold by adding new patches compatible with
${\cal M}_0$. The idea is to construct a chart around each
perturbed state $\rho_X$ as we did around $\rho_0$. Let $\rho_X
\in {\cal M}_0$ with Hamiltonian $H_X \geq I$ and consider the
Banach space ${\cal T}_{(\cdot)}(X)$ of all symmetric forms $Y$
such that the norm $\|Y\|_{(\cdot)}(X)$ is finite, where the
expression for this norm is the same as $\|Y\|_{(\cdot)}(0)$ but
with all the Hamiltonians replaced by $H_X$. Then we repeat
exactly the same process, namely, take sufficiently small $Y$
(with $\|Y\|_{(\cdot)} (X) < 1 - \beta_X$), obtain from the {\em
KLMN} theorem the Hamiltonian $H_{X=Y}$  with form $q_{X+Y} = q_X
+ Y = q_0 + X + Y$ and form domain $Q_{X+Y}=Q_X=Q_0$ and take the
hood of $X$ to be the set ${\cal M}_X$ of all states of the form
$\rho_{X+Y}=Z_{X+Y}^{-1}e^{-H_{X+Y}}=Z_{X+Y}^{-1}e^{-(H_0+X+Y)}$.
The topology and coordinates for ${\cal M}_X$ are then introduced
in a completely similar fashion.

We then turn to the union of ${\cal M}_0$ and ${\cal M}_X$. We
need to show that our two previous charts are compatible in the
overlapping region ${\cal M}_0 \cap {\cal M}_X$. For the case of
form bounded and operator bounded perturbations, the equivalence
of the norms is achieved by a straightforward series of operator
identities \cite{grasselli:Streater98,grasselli:Streater99}. For
the case of $\varepsilon$-bounded perturbations, the argument is
more subtle and involves careful consideration of the domains of
the operators. Nevertheless, the result still follows
\cite{grasselli:GrasselliStreater00}.

\begin{theorem}
$\| \cdot \|_\varepsilon (X)$ and $\| \cdot \|_\varepsilon (0)$
are equivalent norms.
\end{theorem}

We can repeat the construction again, starting from any point in
${\cal M}_0\bigcup {\cal M}_X$. We then obtain the following
definition.

\begin{definition}
The information manifold ${\cal M}(H_0)$ defined by $H_0$ consists
of all states obtainable in a finite numbers of steps, by
extending ${\cal M}_0$ as explained above.
\end{definition}

\section*{Affine Geometry in ${\cal M}(H_0)$}

The set $A=\left\{\widehat X \in \widehat{\cal T}_{(\cdot)} (0):
\widehat X=X-\rho_0 \! \cdot \! X, \|X\|_{(\cdot)}
(0)<1-\beta_0\right\}$ is a convex subset of the Banach space
$\widehat{\cal T}_{(\cdot)}(0)$ and so has an affine structure
coming from its linear structure. We provide ${\cal M}_0$ with an
affine structure induced from $A$ using the patch $\widehat X
\mapsto \rho_X$ and call this the canonical or $(+1)$-affine
structure. The $(+1)$-convex mixture of $\rho_X$ and $\rho_Y$ in
${\cal M}_0$ is then $\rho_{\lambda X+(1-\lambda )Y}$, $(0 \leq
\lambda \leq 1)$ , which differs from the previously defined
$(-1)$-convex mixture \mbox{$\lambda \rho_X +(1-\lambda )\rho_Y$}.

Given two points $\rho_X$ and $\rho_Y$ in ${\cal M}_0$ and their
tangent spaces $\widehat{\cal T}_\varepsilon (X)$ and
$\widehat{\cal T}_\varepsilon (Y)$, we define the $(+1)$-parallel
transport $U_L$ of $(Z-\rho_X \! \cdot \! Z) \in \widehat{\cal
T}_\varepsilon (X)$ along any continuous path $L$ connecting
$\rho_X$ and $\rho_Y$ in the manifold to be the point $(Z-\rho_Y
\! \cdot \! Z) \in \widehat{\cal T}_\varepsilon (Y)$. Clearly
$U_L$ is independent of $L$ by construction, thus the
$(+1)$-affine connection is flat. We see that the $(+1)$-parallel
transport just moves the representative point in the line
$\{Z+\alpha I\}_{\alpha \in {\mathbf R}}$ from one hyperplane to
another.

\section*{Discussion}

The manifold ${\cal M}(H_0)$ constructed here does not cover the
whole set ${\cal M}$ at once. It could not possibly do so, since
our small perturbations do not change the domain of the original
Hamiltonian $H_0$, and certainly ${\cal M}$ contains states
defined by Hamiltonians with plenty of different domains. Also,
although we can reach far removed points with a finite number of
small perturbations, we can not move in arbitrary directions. For
instance, we cannot reach $X=-H_0$ as the result of our
perturbations, since the identity is not an operator of trace
class. To cover ${\cal M}$ entirely, we have to start at several
different points. The whole manifold thus obtained consists of
several disconnected parts, pointing towards positive directions
with respect to the given Hamiltonians.

Finally, it is clear that $\lambda \rho_X +(1-\lambda )\rho_Y$,
for, say, $\rho_X ,\rho_Y \in {\cal M}_0$, defines a new state in
the underlying set ${\cal M}$. Nonetheless, we were not yet able
to prove that it belongs to the neighbourhood of the original
states. This is the main obstacle to define a mixture connection
in our manifold and thence develop a quantum version of Amari's
duality theory \cite{grasselli:Amari85}. One possibility is to
change the definition of the neighbourhoods altogether and use,
for instance, the condition that states in the same neighbourhood
all have finite relative entropy with respect to each other,
besides having finite von Neumann entropy.

\end{document}